\documentclass[12pt]{article}
\usepackage{graphics}
\input epsf

\newcommand{\abs}[1]{\left| #1 \right|}

\newcommand{\be}{\begin{eqnarray}}
\newcommand{\ben}{\begin{eqnarray}\nonumber}
\newcommand{\ee}{\end{eqnarray}}
\newcommand{\ul}{\underline}

\begin{document}
\title{
\vskip 1cm
Metastable Aspects of Singlet Extended Higgs Models
}
\author{L. Clavelli\footnote{lclavell@bama.ua.edu}\\
Department of Physics and Astronomy\\
University of Alabama\\
Tuscaloosa AL 35487\\ }
\maketitle
\begin{abstract}
     It has long been known that the broken supersymmetric (susy) phase of the 
singlet 
extended susy higgs model (SESHM) is at best metastable and the ground
states of the model have vanishing vacuum 
energy and are exactly supersymmetric.  If the SESHM is confirmed at
the Large Hadron Collider (LHC), the numerical values of the parameters
of the model have a bearing on key properties of the susy phase and 
might provide
an estimate of the remaining time before a possible decay of our false vacuum.
We provide some analysis of the model including a treatment of phases
in the potential and soft higgs masses.
\end{abstract}

\section{\bf Introduction}
\setcounter{equation}{0}
    
    In the minimal supersymmetric standard model (MSSM), the higgs sector
consists of two higgs doublets one coupling to the up-type quarks and the
other to the down-type quarks.  This can be extended by the addition of
a higgs singlet and such models have been the subject of extensive study
over the last three decades as chronicled in recent reviews \cite{
Kraml,BLS}.  The most general, renormalizable, singlet extended, susy higgs model
(SESHM) corresponds to the superpotential
\be
    W = \lambda \left( S (H_u\cdot H_d - v^2) + \frac{\lambda^\prime}{3} S^3
           + \frac{\mu_0}{2}S^2 \right) 
\label{superpot}
\ee
where $S$ is the singlet higgs field.  The dot product is defined as
\be
      H_u \cdot H_d = H_u^0 H_d^0 - H_u^+ H_d^- \quad .
\ee
In the present work, focusing on the vacuum properties of the higgs potential, we
suppress the charged higgs fields assuming, as usual, that they do not acquire
vacuum expectation values.

This model was first written by Fayet \cite{Fayet} in the seventies and its
metastability was already noted then.  The vacuum expectation value of the $S$
field corresponds to the $\mu$ parameter of the MSSM and
the introduction of the singlet higgs therefore
provides a possible dynamical solution to the so-called $\mu$ problem of the
MSSM \cite{Nilles}.
Most, if not all, of the subsequent phenomenological
analysis has restricted the most general model by putting one or more of the
parameters in \ref{superpot} to zero \cite{BLS}.  For instance the ``next to minimal
susy standard model" (NMSSM) takes $v$ and $\mu_0$ to zero.  The ``UMSSM"
takes $v$, $\lambda^\prime$, and $\mu_0$ all to zero and introduces an additional
U(1) gauge symmetry.  The ``nearly minimal susy standard model", (nMSSM)
takes $\mu_0$ and $\lambda^\prime$ to zero.  All of these well studied limiting
cases of the SESHM treat the model in the absence of the $\mu_0$ coupling.
The vanishing of $\mu_0$ can be obtained, with some loss of generality, by
requiring that the superpotential is odd under $S\rightarrow -S$ while the
scalar potential is even.
In addition, phenomenological studies have naturally concentrated on the
broken susy phase whereas the properties of the exact susy ground state 
and the inter-phase relations of the model have been largely ignored.  

With the accumulation of experimental evidence that the universe has a 
small positive vacuum energy and originally, in the inflationary era, 
had a much larger vacuum energy, interest in metastable models 
with a susy ground state has increased 
\cite{Giddings,ISS}.  In string theory also, it is thought that the 
universe
makes transitions among a large number of local minimal in a string landscape.
The original and primary manifestations of string theory had zero vacuum 
energy and were exactly supersymmetric as in the ground states of the SESHM.
However, string theory also has a prominent manifestation with negative
vacuum energy, the anti-deSitter universe of the AdS/CFT correspondence, 
and many other possible local minima with a wide variety of vacuum energies.
If the universe were to fall into a state of large negative vacuum energy it
would collapse on a short time scale in a ``big crunch".  This leads to possible
physical and philosophical problems which are irrelevant to this paper since
we treat exclusively the SESHM.  Thus, although string theory can be a source
of inspiration, we do not feel bound by specific string possibilities in 
advance of experimental confirmation.

In the absence of soft susy breaking, the
SESHM has a positive definite scalar potential and, therefore, no negative
vacuum energy solutions.  We assume that the soft breaking terms preserve this
feature in order to provide a model for the current phase of the universe.
Another difference between string theory and lagrangian higgs models is that
in the latter the parameters of the higgs potential are the same in each local
minimum although the vevs of the fields could differ widely.  In string theory
the parameters of the potential are also thought to be minimum dependent.

In a recent paper \cite{Xhiggs} we have shown that the SESHM has four critical
points where all derivatives with respect to the fields vanish.  These are:

\noindent
Solution 1: Exact Susy with Electroweak Symmetry Breaking (EWSB)\\
Solution 2: Exact Susy with no EWSB\\
Solution 3: Broken Susy with no EWSB\\
Solution 4: Broken Susy with EWSB

Solution 4, is, obviously, most close to our universe. 
A susy ground state with EWSB will, in general, support non-zero particle masses 
and therefore susy atoms and molecules \cite{CL}.  Solution 2, on the other hand,
corresponds to a supersymmetric plasma of elementary particles with no atomic, 
molecular, or condensed matter physics. 

\begin{figure}[htbp]
\epsfxsize= 3in 
\leavevmode
\epsfbox{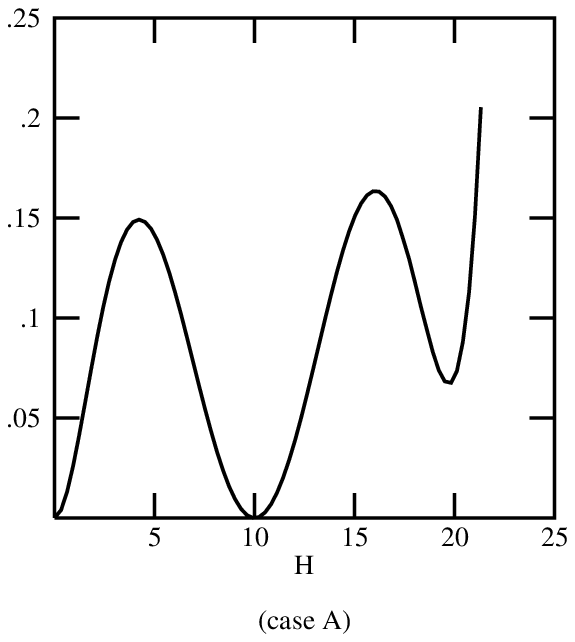}
\epsfxsize= 3in 
\epsfbox{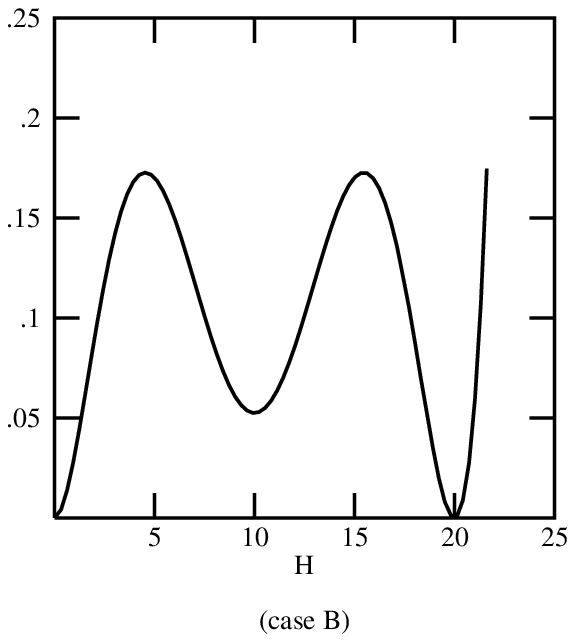}
\caption{The higgs potential is plotted schematically versus a one-dimensional
higgs field in the exothermic case A and the endothermic case B.  The minimum
corresponding to solution 3 is not shown.  See text.}
\label{fig1}
\end{figure}

   We are primarily interested in metastability aspects of the SESHM which 
have not been treated in earlier studies.  It is theoretically possible to identify
which regions of the parameter space have a doublet higgs vacuum expectation
value greater in solution 4 than in solution 1.  Since quark and lepton masses are
controlled by the doublet higgs vev, in matter dominated regions of 
configuration space, these regions of parameter space would
have greater energy in solution 4 than in solution 1.  
This would allow an exothermic
transition from our universe to that of solution 1 whereas, in other regions of
parameter space there would be only an endothermic transition to solution 1
neglecting the small amount of vacuum energy that would be released.
In either case there could be an exothermic transition to solution 2.
The situation is schematically indicated in figure\,\ref{fig1} projected into
a one dimensional higgs space.

Of course, nucleon masses
are proportional to the $\Lambda$ parameter of QCD and are much larger than  
the corresponding quark masses.  Thus, among the atomic constituents, only the lepton masses are sensitive to the higgs vev.  Since in normal matter even the electron energy density
is much greater than the vacuum energy density, the transition to exact susy would 
require extra energy input if the higgs vev in the exact susy minimum were greater than that in our broken susy minimum.  

Although we cannot, a priori, 
predict which region of parameter space is realized in nature, the issue
can be experimentally resolved by a sufficiently detailed study of the higgs 
sector at the LHC or some future accelerator.  We cannot even be sure, of course, that the higgs structure of nature is that of the SESHM but this also will soon be experimentally determined.  Thus the question is one of physics and not merely one of philosophy.  In string theory, on the other hand, it is not clear at present whether there are any unique predictions that are falsifiable in advance of the next transition.

   In ref.\cite{Xhiggs} we treated solution 4 in a simplified toy model ignoring phases and soft susy breaking terms.  In the current article we give the phase analysis. This will eventually allow a treatment of CP violation 
from the higgs sector of
the SESHM although we do not proceed here to a phenomenological analysis.
We also give a partial analysis of soft terms.
  
   There is, at present, no good theory of susy breaking in our universe nor is 
there a good theory for why the vacuum energy is as small as it is.  In the MSSM
and in other phenomenological approaches one usually adds to the scalar potential
explicit ``soft" susy breaking masses and A terms as discussed in section 2.
These are often postulated to derive from susy breaking in some ``hidden" sector
communicated to our sector through gravitational or other interactions
but they could also be thought of as an effective representation of a more
complete model of susy breaking.
In the toy model of ref.\cite{Xhiggs} ignoring phases it was possible to
find consistent higgs sector solutions without treating soft terms.  In the
current treatment of complex fields and complex Lagrangian parameters we find
that it is no longer theoretically consistent to ignore soft terms in the 
broken susy phase.  The current suggestion, then, is that the susy-breaking
critical points of the SESHM are promoted to true minima by non-perturbative 
effects assumed here to be parametrizable in terms of soft higgs masses.  

\noindent
The crucial points are:\\ 
1) When the transition to exact susy occurs, simultaneously
in both sectors, these soft terms will vanish and\\
2) A sufficiently detailed experimental analysis of the broken susy phase can
separate the soft terms from the other terms in the higgs potential and
determine whether the transition to the susy solution 1 is exothermic or not.

A further question that can be answered when all the parameters of the higgs
sector are experimentally determined is what will be the expected lifetime of the 
current universe.  If the higgs vevs in the broken susy state approach those
of the exact susy state or if the height of the barrier goes to zero the 
lifetime of the broken susy phase will go to zero.  
It is, therefore, of interest
to ask for which regions in parameter space, if any, do these conditions prevail.
Experiments at the LHC may thus be relevant to the expected lifetime of the 
current universe.  Of course, the fact that the current universe has survived
for some $13$ billion years also constrains the parameters and suggests that
the expected future lifetime should not be orders of magnitude less than this.  

In section II we review the SESHM and the critical point 
conditions on the parameters and vevs.  A full analysis of the soft terms is postponed
to a time when the LHC has determined that a singlet higgs is, in fact, present.
To illustrate the suggested procedure it is sufficient to consider a soft
higgs scalar mass squared $M_S^2$ and a common soft doublet higgs mass squared
$M_h^2$.  In section III we examine the parameter space as a function of 
these two soft squared masses taken to be positive.  In fact, however, there
are no theoretical or experimental reasons for these soft squared masses to be
positive providing the physical higgs squared masses are non-negative and consistent with experimental constraints.  
 
In section IV we consider the possibility that $M_S^2$ and $M_h^2$ are 
negative which allows
a (fine-tuned) situation in which the higgs vacuum energy in the broken
susy phase is equal to the experimental dark energy.  
We note that there are, at present, no successful models for the dark energy 
which do not involve fine tuning even if the anthropic principle is used to
trivially ``predict" that a theory with a small dark energy exists.
There are also, however,
other possible contributions to the dark energy such as from compactification,
thermal effects, loop contributions with broken susy etc.  Thus, while we find
the possibility of negative $M_S^2$ interesting to consider, its exact 
value cannot be firmly predicted by this method unless or until the other contributions to 
the dark energy are known to be small and the other parameters of the higgs potential are known.

Our results are briefly summarized in section V.

    We rely on the experimental fact that the vacuum energy is 
positive.  We exclude, therefore, contradictory regions of parameter space with 
large negative
values of $M_S^2$ or $M_H^2$.  This has the consequence that the susy minima are
lower in energy density than the broken susy minima and the broken susy minima
are metastable as in ref.\,\cite{Fayet}.  If experiments had 
found a negative vacuum energy in the broken susy phase, it would be possible to
write a SESHM with a broken susy ground state.  These points and the entire 
current paper assume that we can neglect non-perturbative effects except as
effectively describable in terms of the soft terms.  This 
assumption is common to the MSSM and most, if not all, of the phenomenological 
treatments of the SESHM.

\section{\bf The Singlet Extended Susy Higgs Model}
\setcounter{equation}{0}

The superpotential of the SESHM given in eq.\,\ref{superpot} leads to the F term contributions to the scalar potential
\be
   V_F = \lambda^2 \left( \abs{H_u\cdot H_d - v^2 + \lambda^\prime S^2 + \mu_0 S}^2   + \abs{S}^2 (\abs{H_u}^2 + \abs{H_d}^2)\right) \quad .
\label{VF}
\ee  
In addition we have the gauge generated D terms
\be 
     V_D = \frac{g_1^2+g_2^2}{8} \left( \abs{H_d}^2 - \abs{H_u}^2 \right)^2
+ \frac{g_2^2}{2}\left( \abs{H_d}^2 \abs{H_u}^2 - \abs{H_u\cdot H_d}^2\right)^2
\label{Dterms}
\ee
where $g_1$ and $g_2$ are the $U(1)$ and $SU(2)$ gauge coupling constants.
Since we restrict our attention to the potential of the neutral higgs fields, we
can discard the second term in $g_2^2$.

The soft susy breaking terms are
\ben
    V_{soft} &=& \lambda^2 \left( m_{H_u}^2 \abs{H_u}^2 + m_{H_d}^2 \abs{H_d}^2 
+ m_S^2 \abs{S}^2 \right.\\
     &+& \left. (A_s S H_u \cdot H_d  
     + A_1 v S + A_2  \mu_0 S^2 + A_3 \lambda^\prime S^3 + h.c.)\right)\quad .
\label{Vsoft}
\ee

The full potential is the sum of these three terms
\be
    V = V_F + V_D + V_{soft}\quad .
\ee

Since the soft parameters are, a priori, undetermined we can, without loss of
generality, factor out a $\lambda^2$ as in eq. \,\ref{Vsoft}.  A result of this
convention is that the critical point conditions do not depend on $\lambda$
although all the mass squared eigenstates are proportional to $\lambda^2$ apart 
from small calculable D term contributions.

The critical points of the potential are those points in field space where all
first derivatives with respect to the fields vanish.  The critical points are true
local minima if the eigenvalues of the higgs mass squared matrix are all non-negative.
  
In the exact susy phase, the soft terms are absent as discussed in the introduction.
The symmetry of the remaining potential under $H_u - H_d$ interchange, implies that
at the minima of the potential the two doublet higgs will have equal vacuum expectation
values and the D terms will vanish which is, in fact, a condition for exact susy.

   The most general renormalizable model including soft terms corresponds to a 
complicated multi-parameter potential whose full analysis will only be justified when and if singlet higgs fields are experimentally discovered. In this paper we  show how certain questions related to metastability can be addressed while restricting our
attention to soft mass terms only, with the additional simplification of equal
soft $H_u$ and $H_d$ mass squared terms.  This latter restriction leads to equal
doublet higgs vevs in the susy breaking minima and therefore to the absence of
D terms at the minima. The D terms will however contribute to the mass squared 
matrix and must therefore be taken into account.

We analyse, therefore, the simplified model
\be
    V = V_F + V_D + \lambda^2 m_S^2 |S|^2 + \lambda^2 m_H^2 (|H_u|^2 + |H_d|^2)\quad .
\label{scalarpot}
\ee 

Hermiticity requires that $\lambda$, $m_S^2$, and $m_H^2$ be real.
Without loss of generality we can take one further parameter real; we choose $v$ real.  The phase dependence of the potential then corresponds only to the phases of the remaining three terms in $V_F$ so
the phase of $\lambda^\prime$ can be absorbed into that of the field
$S$ and the parameter $\mu_0$, $\lambda^\prime$ then being taken real.
Apart from the overall
factor of $\lambda^2$ the Lagrangian is then dependent on the 
the complex parameter $\mu_0$ and on the real parameters $m_S^2$,
$m_H^2$, $\lambda^\prime$ and $v$,
a total of six real parameters. 
As a function of these we seek points in the field space 
\ben
   <S>&=&S_0\\ 
   <H_u>&=&<H_d>=v_0 
\ee
at which all the first derivatives of V vanish.
The critical point conditions are, therefore,
\be
    \frac{1}{\lambda^2} \frac{\partial V}{\partial S^\dagger}|_{_0} = 0 =
  (2 \lambda^\prime S_0^* + \mu_0^*)(v_0^2 - v^2 + \lambda^\prime S_0^2 
  + \mu_0 S_0) + S_0 (2 |v_0|^2 + m_S^2)  
\label{CP1}
\ee
and
\be
    \frac{1}{\lambda^2} \frac{\partial V}{\partial H_u^\dagger}|_{_0} = 0 =
    v_0^* (v_0^2 - v^2 + \lambda^\prime S_0^2 + \mu_0 S_0) + v_0 (|S_0|^2 + m_H^2)\quad .
\label{CP2}
\ee

These two equations determine possible values of the complex vevs, $S_0$ and $v_0$ in terms
of the parameters of the potential.  It is convenient, however, to invert the relations
and consider $\mu_0$ and $v$ as determined by the vevs $S_0$ and $v_0$.

The higgs contribution to the vacuum energy is
\be
    V(0) = \lambda^2 \left(|v_0^2 - v^2 + \lambda^\prime S_0^2 + \mu_0 S_0|^2
  + 2|S_0|^2|v_0|^2 + m_S^2 |S_0|^2 + 2 |v_0|^2 m_H^2 \right ) \quad .
\label{vacen}
\ee 

The critical points are independent of $\lambda$ while the higgs vacuum energy is proportional to $\lambda^2$ but this cannot be made
arbitrarily small without similarly reducing the masses of the physical higgs eigenstates unless the soft masses $m_S^2$ and $m_H^2$ are correspondingly fine-tuned. We will set $\lambda$ to unity with the understanding that the physical higgs masses can be scaled by a common factor of $\lambda$ while the vacuum energy density is scaled by a factor of $\lambda^4$.

The critical points with {\underline exact} susy are found by solving eqs.\,\ref{CP1} and \ref{CP2} neglecting the soft mass terms as discussed in the introduction.  Two degenerate solutions are found with zero vacuum energy:

\noindent
Solution 1:  $v_0 = v\quad$,$\quad S_0=0\quad$: exact susy with EWSB\\
Solution 2:  $v_0 = 0\quad$,$\quad \lambda^\prime S_0^2 + \mu_0 S_0 -v^2 = 0\quad$: exact susy; no EWSB

Since, neglecting soft terms, the scalar potential is positive definite or zero, any localized critical points with zero vacuum energy are true minima.  

Two more solutions can be found with non-zero vacuum energy
and non-zero singlet vev $S_0$.  For these, to illustrate the procedure, we keep the soft mass terms as representative of all the soft terms .
If a singlet higgs is confirmed at LHC it will be of interest to do the complete global analysis.

Solution 3 has no EWSB ($v_0=0$) and a cubic equation for $|S_0|$:
\be
    (2 {\lambda^\prime} S_0^* + \mu_0^*)(-v^2 + \lambda^\prime S_0^2 + \mu_0 S_0)+S_0 m_S^2=0\quad .
\ee
For this to be a true local minimum we would have to require that all the 
physical higgs squared masses are non-negative here.  We will not pursue this question at solution 3 in the present paper
although it might ultimately become interesting to consider whether a transition from
solution 3 to solution 4 could describe an EWSB transition in the early universe.
Here we concentrate on the parameter space conditions that will make solution 4
a true local minimum.  Solution 4, which has EWSB, corresponds to the extensive phenomenological discussion in the literature \cite{Kraml,BLS} treating various limiting cases of the SESHM.
In solution 4, we take the experimental EWSB to require $|v_0| \approx 247\,$GeV.  

With non-zero $v_0$ and $S_0$, the two equations \ref{CP1} and \ref{CP2} can be combined to yield
\be
      - \frac{v_0^*}{v_0}(2 \lambda^\prime S_0 + \mu_0)(|S_0|^2+m_H^2) + S_0^* (2|v_0|^2+m_S^2) = 0
\label{cond1}
\ee
and
\be
  v^2 = {v^*}^2=v_0^2 + \lambda^\prime S_0^2 + \mu_0 S_0 + \frac{v_0}{v_0^*}(|S_0|^2 + m_H^2)
\quad .
\label{cond2}
\ee
The vacuum energy in solution 4, $V_4(0)$, is given by substituting the vevs $v_0$ and
$S_0$ satisfying eqs.\,\ref{cond1} and \ref{cond2} into eq.\,\ref{vacen}.
\be
   V_4(0)= \lambda^2 \left( |S_0|^4 + |S_0|^2 (2|v_0|^2 + m_S^2 + 2 m_H^2) + m_H^4 + 2 m_H^2 |v_0|^2 \right) \quad .
\label{V4(0)}
\ee

It is convenient to define the real dimensionless variables
\be
    A = |v_0/S_0|\quad ,
\ee
\be
    B = (2|v_0|^2+m_S^2)/|S_0|^2 \quad ,
\ee
and
\be
    C = 1 + m_H^2/|S_0|^2 \quad.
\ee
Rather than deal immediately with the cubic equation for $S_0$, we prefer to
define the parameter $\mu_0$ in terms of the value for $S_0$ assumed to run over
all consistent values.  From eqs.\,\ref{cond1} and \ref{cond2}:
\be
    \mu_0 = \frac{S_0^* B v_0}{v_0^* C} - 2 \lambda^\prime S_0 \quad .
\label{mu0}
\ee
We could, of course, investigate the $\mu_0=0$ case as in the well-studied limiting cases of the SESHM but we prefer to maintain generality.

\section{\bf The Higgs Mass Squared Matrix}
\setcounter{equation}{0}

The higgs mass squared matrix in solution 4 is obtained by taking the
scalar potential of eq.\,\ref{scalarpot} and expanding around the
vevs of the neutral fields.  We write
\ben
     S &=& s + S_0\\
     H_u &=& h_u + v_0\\\nonumber
     H_d &=& h_d + v_0 
\ee

Taking $v$ real, we can write the four complex numbers as
\ben
    \mu_0 &=&|\mu_0| e^{i \phi_\mu}\\\nonumber
    S_0 &=& |S_0|e^{i\phi}\\
    v_0 &=& |v_0|e^{i\phi_0} \quad .
\ee
It is convenient to use the symmetric and antisymmetric combinations of the
$h$ fields:
\ben
     h_u &=& h_p + h_m\\
     h_d &=& h_p - h_m \quad .
\ee
Using eqs.\,\ref{cond1} and \ref{cond2} to eliminate the terms linear in
the shifted fields and discarding terms cubic and quartic in the fields, we
have
\be
    V = V_4(0)+ X \quad .
\ee
The mass squared matrix can be written
\be
    X = \Psi_1^2 M_1^2/2 + \Psi_2^2 M_2^2/2 + \Delta X \quad .
\ee
The two real fields $\Psi_1$ and $\Psi_2$ have been trivially separated out with
the definitions
\ben
    \Psi_1 &=& h_m e^{-i\phi_0} + h_m^\dagger e^{i\phi_0}\\
    \Psi_2 &=& h_m e^{-i\phi_0}/i - h_m^\dagger e^{+i\phi_0}/i 
\ee
and the corresponding squared masses
\ben
     M_1^2 &=& 2 C |S_0|^2 + 8 (g_1^2+g_2^2) |v_0|^2\\
     M_2^2 &=& 0 \quad .
\ee
The massless field is absorbed into a massive Z boson.  The remaining 
scalar boson can be
quite light depending on the values of $C$ and $|S_0|^2$.
The possibility of observing such a light higgs boson of the NMSSM in
the radiative decay of the $\Upsilon$ has been recently discussed \cite{Dermisek}.

The remaining mass squared matrix is
%
%
\ben
\Delta X &=&
    -h_p^2 C |S_0|^2 v_0^*/v_0
    +2 h_p h_p^\dagger (2|v_0|^2+C |S_0|^2)
    +2 h_p s S_0^* v_0^*\\\nonumber
   && +2 h_p s^\dagger S_0 v_0^* (B/C+ 1)
    -{h_p^\dagger}^2 C |S_0|^2 v_0/v_0^*
    +2 h_p^\dagger s S_0^* v_0 (B/C+1)
    +2 h_p^\dagger s^\dagger S_0 v_0\\
   && -s^2 C |S_0|^2 \lambda^\prime v_0^*/v_0
    +s s^\dagger |S_0|^2 B(B/C^2+1)
    -{s^\dagger}^2 C |S_0|^2 {\lambda^\prime} v_0/v_0^* \quad .
\ee
This can be simplified
and written in terms of real fields, $\Phi_i$, by the transformations:
\ben
     h_p &=& (\Phi_1 + i \Phi_2)e^{i\phi_0}\\
     h_p^\dagger &=& (\Phi_1 -i \Phi_2)e^{-i\phi_0}\\\nonumber
     s &=& (\Phi_3 + i \Phi_4) e^{i\phi_0} \\
     s^\dagger&=& (\Phi_3 -i \Phi_4) e^{-i\phi_0} \quad .
\ee
Then
\ben
   \Delta X &=& 4 \Phi_1^2 |v_0|^2 + 4 \Phi_2^2 (|v_0|^2 + C |S_0|^2)
   + \Phi_3^2 |S_0|^2 (B^2/C^2 + B - 2 C |\lambda^\prime|\cos(\phi_\lambda))\\\nonumber
  && + \Phi_4^2 |S_0|^2 ( B^2/C^2 + B + 2 C \lambda^\prime\\\nonumber
  && + 4 \Phi_1 \Phi_3 |S_0|^2 A \cos(\phi-\phi_0) (B/C+2)
     + 4 \Phi_1 \Phi_4 |S_0|^2 A \sin(\phi-\phi_0) (B/C+2)\\
  &&  - 4 \Phi_2\Phi_3 |S_0|^2 A B \sin(\phi-\phi_0) 
      + 4 \Phi_2 \Phi_4 |S_0|^2 A B \cos(\phi-\phi_0) \quad .
\ee
Next we make the transformations:
\ben
   \Phi_3&=&X_3 \cos(\phi-\phi_0) - X_4 \sin(\phi-\phi_0)\\\nonumber
   \Phi_4&=&X_3 \sin(\phi-\phi_0) + X_3 \cos(\phi-\phi_0)\\\nonumber
   \Phi_1&=&X_1\\
   \Phi_2&=&X_2
\ee
after which
\ben
 \Delta X &=& 4 X_1 X_1 |v_0|^2 +4 X_1 X_3 |v_0|^2 (B/C+2)
    + 4 X_2 X_2 |v_0|^2 (1+C/A^2) +4 X_2 X_4 |v_0|^2 B/C \\\nonumber
    &+&X_3 X_3 |S_0|^2 (B^2/C^2+B-2 C \lambda^\prime\cos(\beta) )
    +4 X_3 X_4 |S_0|^2 C \lambda^\prime \sin(\beta)\\
    &+&X_4 X_4 |S_0|^2 (B^2/C^2+B+2 C \lambda^\prime\cos(\beta)) \quad .
\ee
%
%
where we have put
\be
    \beta= 2 (\phi-\phi_0) \quad  . 
\ee

We scan over six real parameters, $\lambda^\prime$, $m_S^2$, $m_H^2$, $|S_0|$,$v$, 
and $\phi$. 
If $\lambda^\prime$ vanishes as in the nMSSM, it is possible to diagonalize the
mass matrix analytically.  We prefer, however, to maintain greater generality so
we numerically solve for the eigenvectors and the mass squared eigenvalues using 
the Jacobi method \cite{numrecipes}.  We require that all the eigenvalues are
non-negative and examine the resulting solution space.  In the numerical study
we take $|v_0|$, nominally $247$ GeV, as our unit of energy.  In these units the 
exothermic situation discussed above then requires that $v$ be less than unity.

From eq.\,\ref{mu0} we have
\be
    |\mu_0| = |S_0| \left( B^2/C^2 + 4 {\lambda^\prime}^2 - 4 \lambda^\prime (B/C)\cos(\beta)\right)^{1/2} \quad .
\ee
From eqs.\,\ref{cond1},\ref{cond2},and \ref{mu0}
\be
    v^2 = e^{2i\phi_0} \left( |v_0|^2 - \lambda^\prime |S_0|^2 e^{i\beta} 
+ |S_0|^2(B/C+C) \right) \quad .
\ee
The condition that $v$ is real determines the phase $\phi_0$ in terms of five
real parameters:
\be
   \tan(2 \phi_0) = \frac{\lambda^\prime|S_0|^2 \sin(\beta)}
{|v_0|^2 + |S_0|^2(B/C+C)-\lambda^\prime |S_0|^2 \cos(\beta)} 
\ee
after which
\be
     v^2 = \cos(2 \phi_0)(|v_0|^2 + |S_0|^2 (B/C+C)) - \lambda^\prime |S_0|^2 \cos(2 \phi_0 + \beta)
\ee 
and
\be
    \phi = \beta/2 + \phi_0 \quad .
\ee

The phase of $\mu_0$ is determined by the reality of $v^2$:
\be
     \tan(\phi_\mu) = \frac{\tan(2 \phi_0 - \phi)(B/C - 2|\lambda^\prime|\cos(\beta)}
{B/C - 2 \lambda^\prime (\cos(\beta) + \sin(\beta) \tan(2 \phi_0 - \phi)} \quad .
\ee 

We now scan over the six free parameters requiring that the squared higgs masses
be positive and recording the consistent values of the parameters.  

\section{\bf Solutions with positive soft squared masses}
\setcounter{equation}{0}

\begin{figure}[htbp]
\begin{center}
\epsfxsize= 5.5in 
\leavevmode
\epsfbox{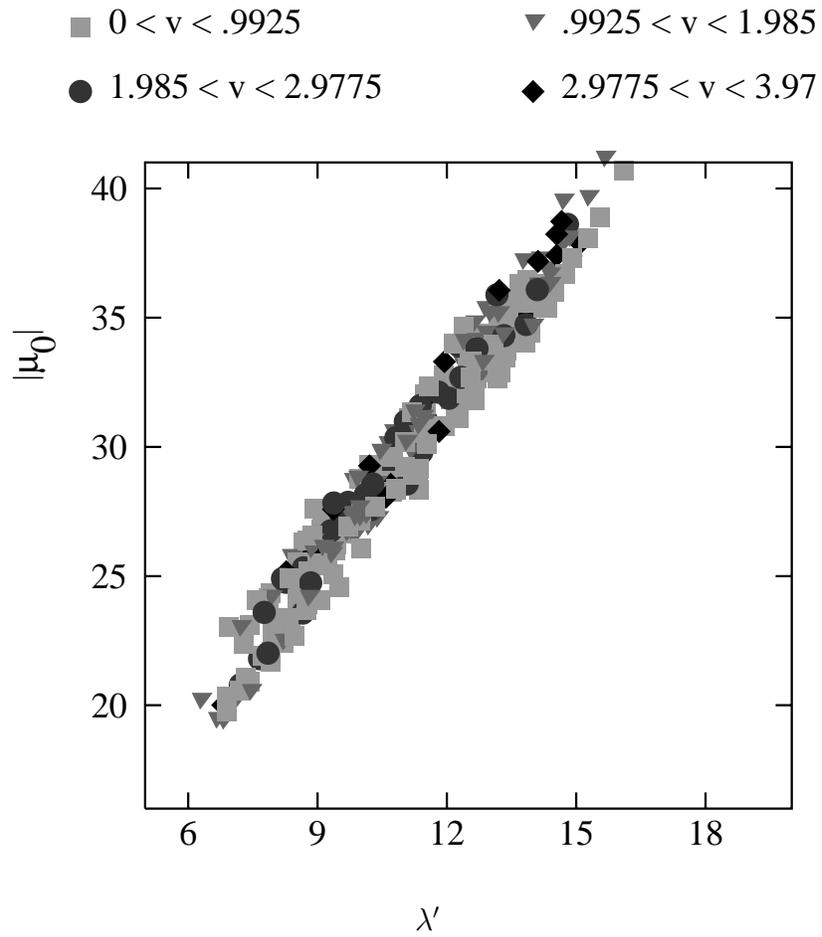}
\end{center}
\vskip -1.5cm

\centerline{$\lambda^\prime$}

\caption{The space of solutions for $\lambda^\prime$ and $|\mu_0|$ with the corresponding-
values of $v$ grey-scale coded in the case of positively constrained soft squared masses.}
\label{elmuv12}
\end{figure}

\begin{figure}[htbp]
\begin{center}
\epsfxsize= 5.5in 
\leavevmode
\epsfbox{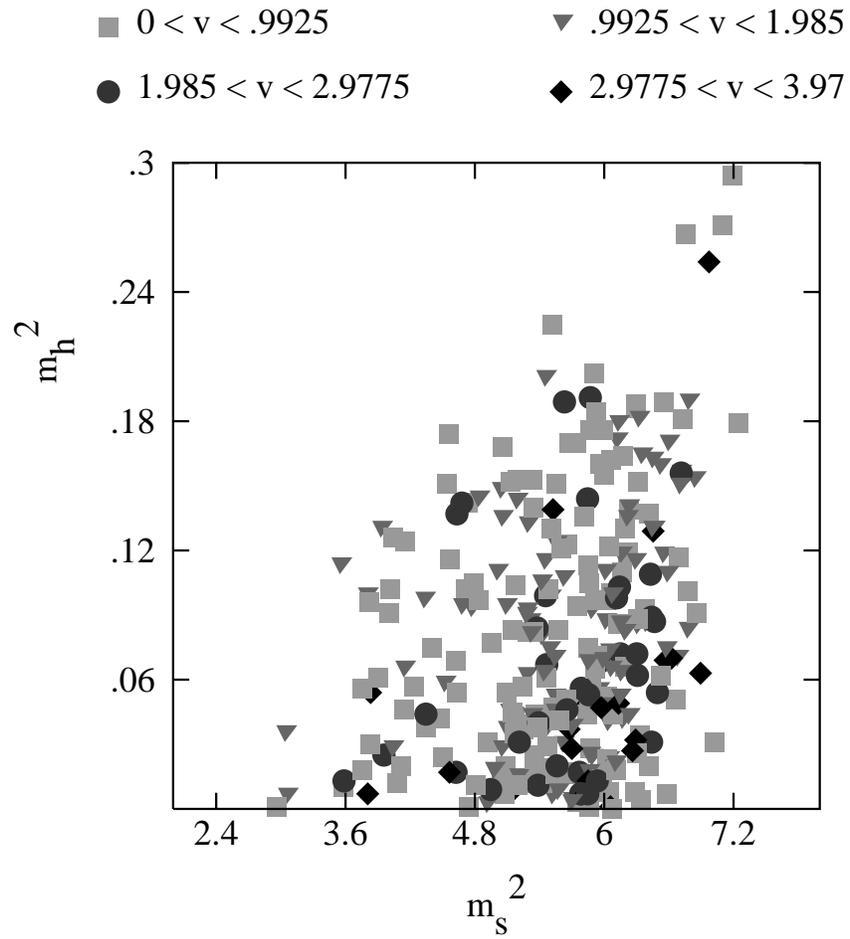}
\end{center}
\vskip -1cm

\caption{The space of solutions for positive soft squared masses $m_s^2$ and $m_h^2$ with the corresponding
values of $v$ grey-scale coded.}
\label{msmhv12c}
\end{figure}

\begin{table}[htbp]
\begin{center}
\begin{tabular}{||c|c|c|c|c||}\hline
   &          &           &        &           \\
   &  minimum & maximum   &  mean  &  std dev  \\
\hline
\hline
                 &                &              &         &          \\
$\lambda^\prime$ &  $5.93$        & $16.1$       &  $11.0$ &   $2.19$  \\
$M_3$            &  \ul{$0.32$}   & $1.17$       &  $0.55$ &   $0.15$  \\
$M_4$            &  \ul{$0.32$}   & $4.09$       &  $1.34$ &   $0.46$  \\
$M_5$            &  $6.19$        & \ul{$10.0$}  &  $9.06$ &   $0.76$  \\
$M_6$            &  $0.34$        & $7.98$       &  $5.80$ &   $0.76$  \\
$V_4(0)$         &  $3.0$         & $4.0 $       &  $3.7 $ &   $0.21$  \\
$|S_0|$          &  $0.96$        & $1.11$       &  $1.05$ &   $0.03$  \\
$M_S^2$          &  $2.37$        & \ul{$7.27$}  &  $5.53$ &   $0.84$  \\
$M_H^2$          &  $0.0$         & $0.29$       &  $.074$ &   $.058$  \\
$v$              &  \ul{$0.0$}    & \ul{$4.0 $}  &  $1.2 $ &   $0.90$  \\
$|\mu_0|$        &  $16.9$        & $41.0$       &  $29.8$ &   $4.75$  \\
$\phi$           &  $0.0$         & $6.283$      &  $3.14$ &   $1.85$  \\
$\phi_0$         &  $2.36$        & $5.49$       &  $4.03$ &   $0.88$  \\
$\phi_\mu$       &  $0.00$        & $6.283$      &  $3.11$ &   $2.54$  \\
$\beta$          &  $2.75$        & $3.52$       &  $3.14$ &   $.097$  \\
$M_{min}$        &  \ul{$0.32$}   & $0.94$       &  $0.52$ &   $0.14$  \\
                 &                &              &         &           \\
\hline
\end{tabular}
\end{center}
\caption{Allowed values of parameters requiring positive physical higgs squared masses and positive soft squared
masses in solution 4.  $M_3$ through $M_6$ together with $M_1$ give the range of 
physical higgs masses.  Underlined quantities are somewhat arbitrarily imposed.  
Other quantities are
consequent limits of the solution space.  $M_{min}$ gives the minimum value
of $M_1$ and $M_3$ through $M_6$.  Dimensional quantities are
given in terms of the doublet higgs vev $v_0=247$ GeV. 
}
\label{pos}
\end{table}

In this section we constrain the soft squared masses to be positive.  We scan
for solutions with
the physical higgs squared masses between 0.1 and 10 in units of $|v_0|^2$,
$0<v<4$, and soft squared masses between $0$ and $7.5$.
This first requirement is equivalent to the assumption that the minimum
higgs mass is greater than about $80$ GeV although, as noted above and in ref.\,\ref{Dermisek}, in the
SESHM, the experimental limits need to be carefully re-examined.  
As noted in the introduction, no solutions are found with both soft masses vanishing
although copious solutions are found with vanishing $m_H^2$.  The vacuum energy
density at the minimum lies between $2.9$ and $4.0$ in units of $|v_0|^4$.  This
agrees with the expectation that the vacuum energy is of order of the higgs vev
to the fourth power.  It is of course far greater than the observed value of
the vacuum energy in our broken susy state.
 
A distinctive property of this solution space is that $\lambda^\prime$ and
$|\mu_0|$ are in the non-perturbative region.  This is similar to the finding
of large couplings in another model of the metastable vacuum \cite{ISS}. 

\section{\bf Solutions with negative soft squared masses}
\setcounter{equation}{0}

\begin{figure}[htbp]
\begin{center}
\epsfxsize= 5.5in 
\leavevmode
\epsfbox{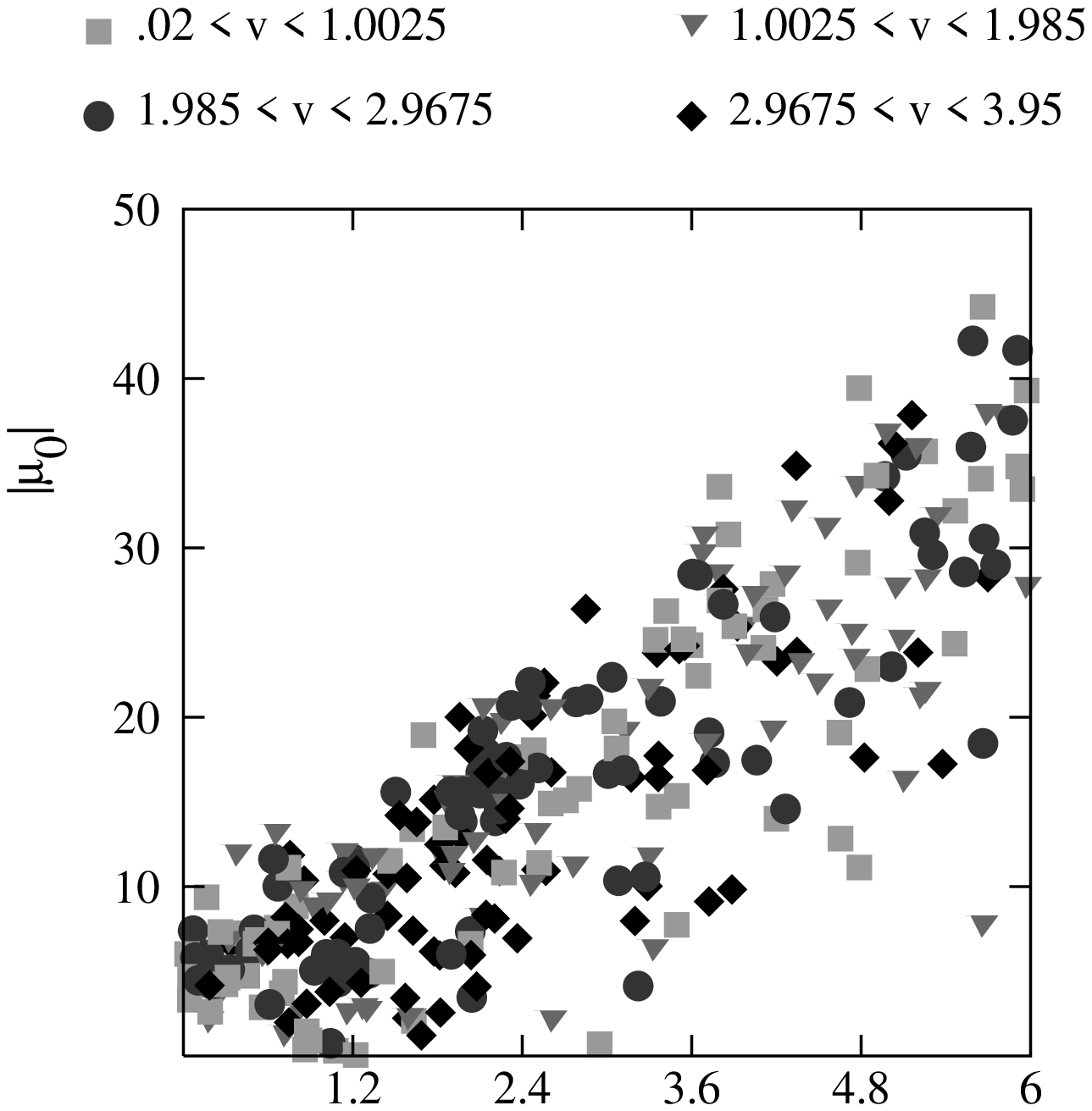}
\end{center}
\vskip -1.5cm

\centerline{$\lambda^\prime$}

\caption{The space of solutions for $\lambda^\prime$ and $|\mu_0|$ with the corresponding
values of $v$ grey-scale coded allowing for negative squared soft masses and requiring that the higgs vacuum energy be equal to the
observed value of dark energy.}
\label{elmuv13}
\end{figure}

\begin{figure}[htbp]
\begin{center}
\epsfxsize= 5.5in 
\leavevmode
\epsfbox{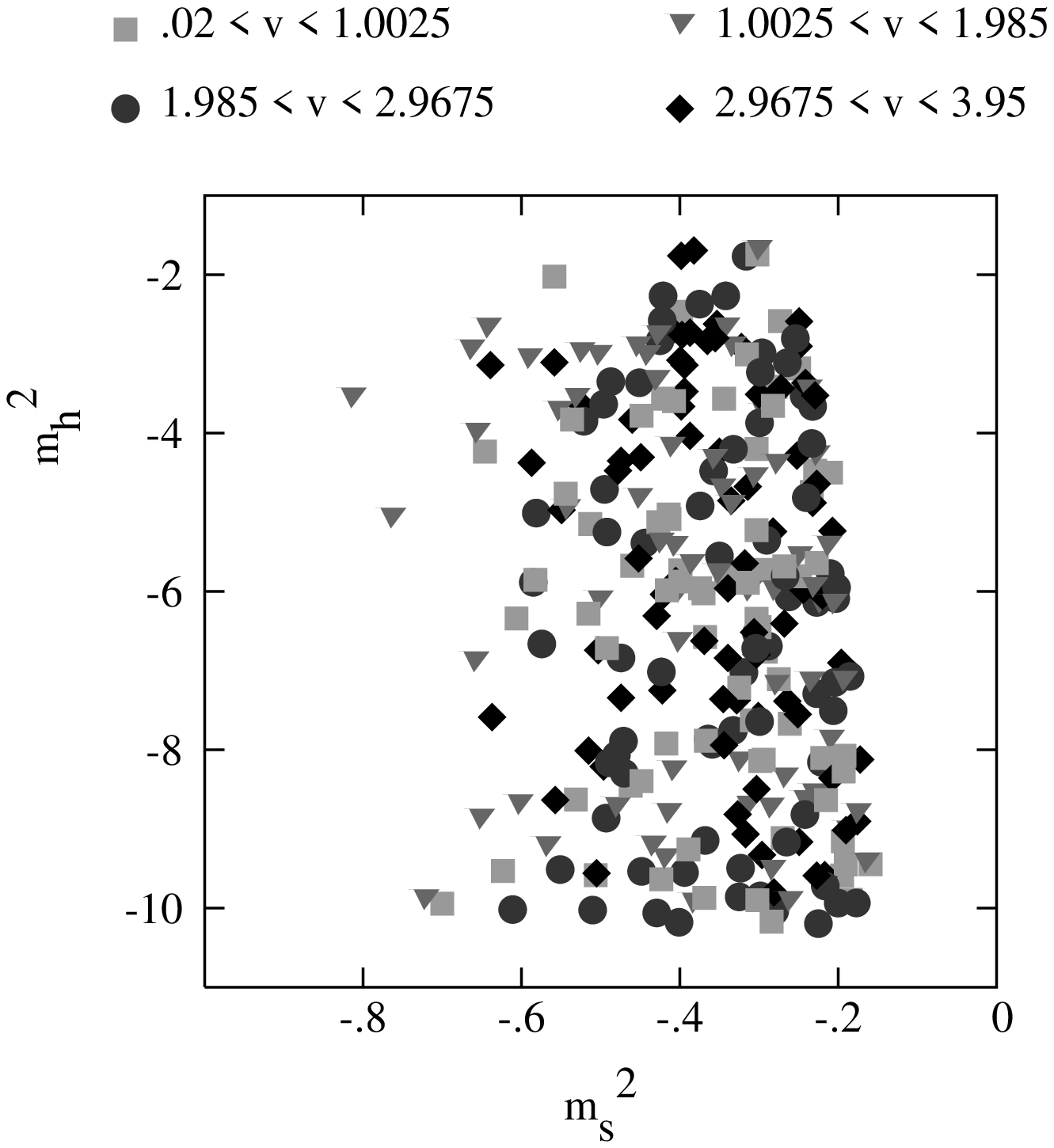}
\end{center}
\vskip -1cm

\caption{The space of solutions for $m_s^2$ and $m_h^2$ is shown with the corresponding
values of $v$ grey-scale coded allowing for negative squared soft masses and requiring that the higgs vacuum energy be equal to the
observed dark energy.}
\label{msmhv13}
\end{figure}

\begin{figure}[htbp]
\begin{center}
\epsfxsize= 5.5in 
\leavevmode
\epsfbox{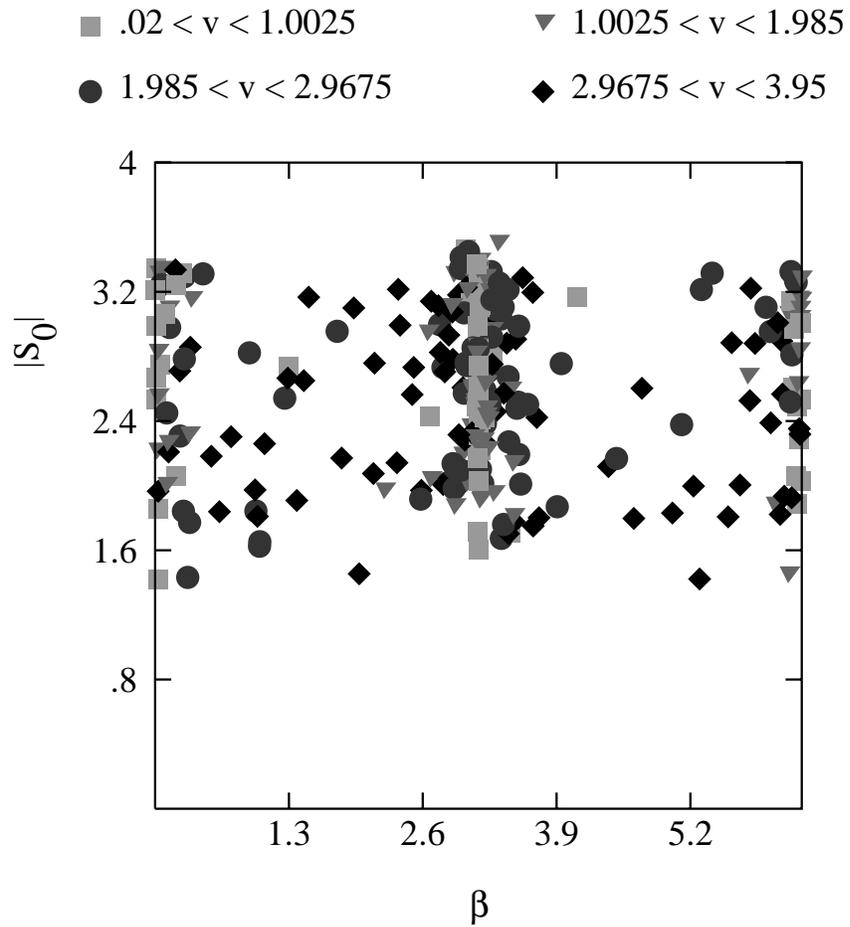}
\end{center}
\vskip -1cm

\caption{The space of solutions for $\beta$ and $|S_0|$ with the corresponding
values of $v$ grey-scale coded allowing for negative squared soft masses and requiring that the higgs vacuum energy be equal to the
observed dark energy.  It is seen that the low values of $v$ are concentrated
at $\beta=n \pi$ }
\label{beS0v13}
\end{figure}

\begin{figure}[htbp]
\begin{center}
\epsfxsize= 5.5in 
\leavevmode
\epsfbox{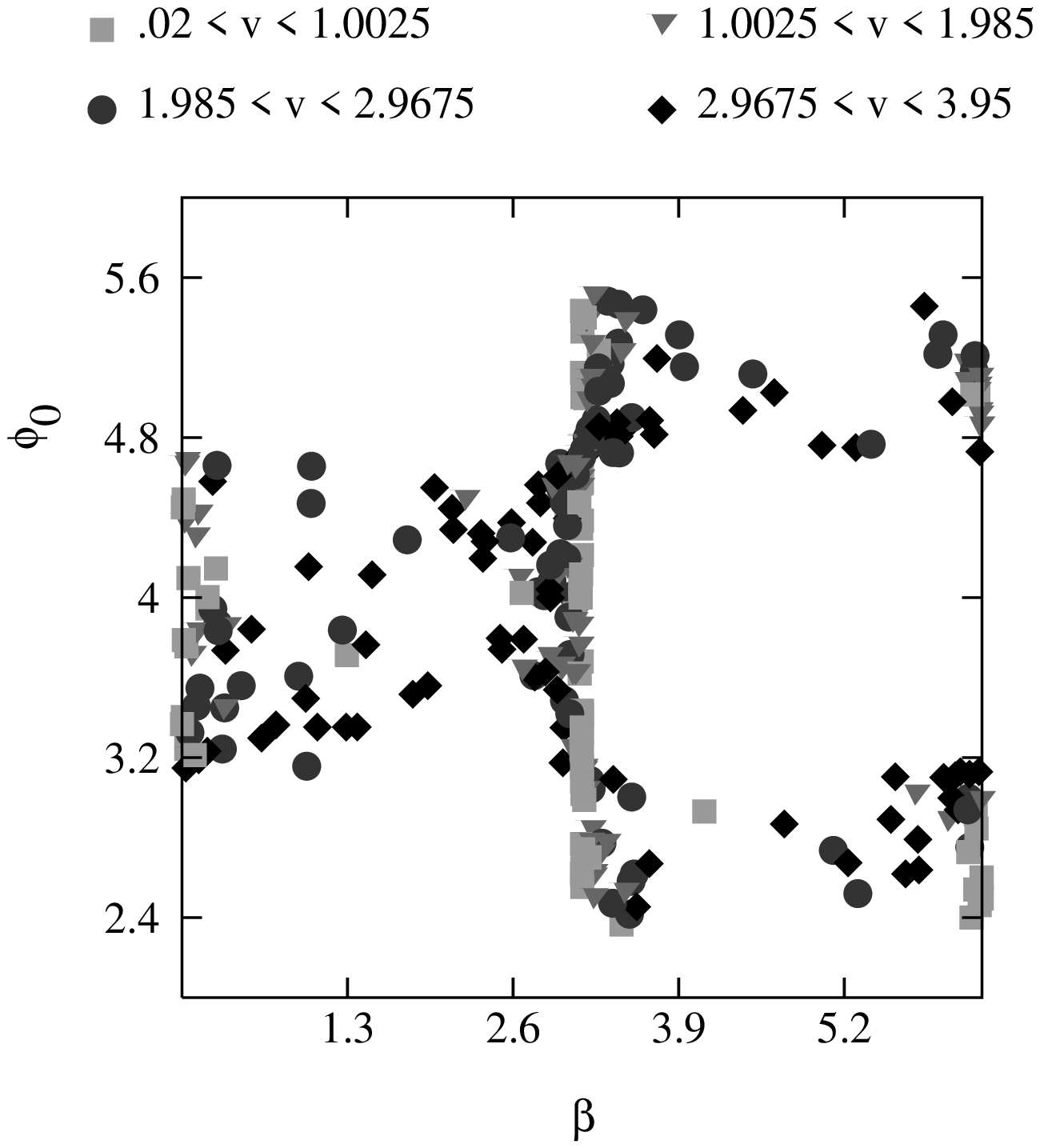}
\end{center}
\vskip -1cm

\caption{The space of solutions for $\beta$ and $\phi_0$ with the corresponding
values of $v$ grey-scale coded allowing for negative squared soft masses and requiring that the higgs vacuum energy be equal to the
observed dark energy.}  
\label{betaphic}
\end{figure}

\begin{table}[htbp]
\begin{center}
\begin{tabular}{||c|c|c|c|c||}\hline
   &          &           &        &           \\
   &  minimum & maximum   &  mean  &  std dev  \\
\hline
\hline
                 &                &              &         &          \\
$\lambda^\prime$ &  $0.010$       & $5.99$       &  $2.70$ &   $1.66$  \\
$M_3$            &  \ul{$0.33$}   & $1.72$       &  $1.23$ &   $0.33$  \\
$M_4$            &  $1.58$        & $3.55$       &  $2.59$ &   $0.37$  \\
$M_5$            &  $3.34$        & \ul{$9.93$}  &  $6.11$ &   $1.57$  \\
$M_6$            &  \ul{$0.327$}  & \ul{$9.75$}  &  $5.44$ &   $2.21$  \\
$V_4(0)$  & \ul{$1.55 \cdot 10^{-57}$}  & \ul{$1.63 \cdot 10^{-57}$}  & \ul{$1.59 \cdot 10^{-57}$} & \ul{$0.04 \cdot 10^{-57}$}    \\
$|S_0|$          &  $1.37$        & $3.48$       &  $2.64$  &   $0.52$  \\
$M_S^2$          &  $-0.85$       & $-0.16$      &  $-0.37$ &   $0.13$  \\
$M_H^2$          &  $-10.2$       & $-1.62$      &  $-6.38$ &   $2.40$  \\
$v$              &  \ul{$0.0$}    & \ul{$4.0 $}  &  $2.01$  &   $1.14$  \\
$|\mu_0|$        &  $0.05$        & $45.2$       &  $15.9$  &   $10.2$  \\
$\phi$           &  $0.0$         & $6.283$      &  $3.20$  &   $1.76$  \\
$\phi_0$         &  $2.36$        & $5.48$       &  $3.96$  &   $0.88$  \\
$\phi_\mu$       &  $0.00$        & $6.283$      &  $3.21$  &   $2.29$  \\
$\beta$          &  $0.00$        & $6.283$      &  $3.08$  &   $1.76$  \\
$M_{min}$        &  \ul{$0.33$}   & $1.61$       &  $1.09$  &   $0.26$  \\
$C$              &  $-0.06$       & $0.26$       &  $0.13$  &   $0.04$  \\       
                 &                &              &          &           \\
\hline
\end{tabular}
\end{center}
\caption{Allowed values of parameters requiring that the higgs vacuum energy
in solution 4 be equated with the observed vacuum energy.  This requires that
both of the soft squared masses are negative although the physical higgs
squared masses are all required to be positive. 
$M_3$ through $M_6$ together with $M_1$ give the range of 
physical higgs masses.  Underlined quantities are imposed.  Other quantities are
consequent limits of the solution space.  $M_{min}$ gives the minimum value
of $M_1$ and $M_3$ through $M_6$.
}
\label{neg}
\end{table}

    As discussed in the introduction, the soft squared masses can be negative as long
as the physical higgs particles have positive squared masses, a condition we can
impose in the scan over parameters.  The freedom to consider negative soft squared
masses can be used to parameterize the fine tuning of the vacuum energy that is,
at present, necessary to understand its smallness relative to the natural scale
of $v_0^4$.  Ignoring other possible contributions we may equate $V_4(0)$ with
the observed vacuum energy \cite{PDG}:
\be
    V_4(0) = (5.9\pm 0.2 ) {\displaystyle meV}^4 = (1.59 \pm 0.04)\cdot 10^{-57} |v_0|^4
\ee 
We may fix $V_4(0)$ at this value by choosing in eq.\,\ref{V4(0)}
\be
     m_S^2 = - C^2 |S_0|^2 - 2 |v_0|^2 C + |v_0|^4 \frac{1}{\lambda^2 |S_0|^2}\cdot 10^{-57} 
  (1.59 + 0.04 (2 r - 1)) 
\ee
where $r$ is a random number between $0$ and $1$.
It is interesting that in this part of parameter space it is possible to find solutions with small values of the couplings $\lambda^\prime$ and $|\mu_0|$  thus suggesting a perturbative treatment of the higgs interactions.  
Some scatter plots of these solutions are shown in figs.\,\ref{elmuv13},\ref{msmhv13},\ref{beS0v13}, and
\ref{betaphic}.  It can be noted that, in this full
theory with complex fields, no solutions are found for the phase of the doublet higgs vev, $\phi_0$, near $0$ or $2 \pi$.  The hypercube containing all the solutions
is tabulated in table\,\ref{neg}.

\section{\bf Summary}
\setcounter{equation}{0}

We have examined the singlet extended higgs model from the point of view of a 
possible transition from our world to an exactly supersymmetric world with electroweak symmetry breaking.  We have included the effects of the quadratic superpotential term in the scalar higgs field
which has been neglected in the analyses of other authors.  
Copious parameter space solutions have been found in which an exothermic transition to an exact susy world with EWSB is possible as well as many others where this is not possible.  Experiments at the LHC will be able to decide which solutions are realized in nature if, in fact, a susy higgs structure with an extra singlet is confirmed.
We have begun the analysis of phases in the potential which could eventually be of interest for CP violation.  We have noted and explored physical regions of parameter space where the soft squared higgs masses are negative although the physical higgs squared masses are positive.

In a future paper we would like to explore solution 3 which has exact susy but no EWSB. It will be of interest to ask whether some or all of the physical regions of parameter space for solution 4 will also lead to a true minimum of the potential at the critical point of solution 3.
In this case we will be able to consider whether the model is consistent with an EWSB transition from solution 3 to solution 4 occuring in the early universe. 

Phenomenological predictions of the model for production mechanisms and
decay chains at the LHC is left for consideration in later papers
with special attention to the effects of a non-zero $\mu_0$ parameter.

{\bf Acknowledgements}

    This work was supported in part by the US Department of Energy under
grant DE-FG02-96ER-40967.

\end{document}